\documentstyle[12pt,epsf]{article}
\newcommand{\nc}{\newcommand}
\def\frac#1#2{{\textstyle {#1 \over #2}}}

\nc{\beq}{\begin{equation}}
\nc{\eeq}{\end{equation}}
\nc{\beqa}{\begin{eqnarray}}
\nc{\eeqa}{\end{eqnarray}}
\nc{\lsim}{\begin{array}{c}\,\sim\vspace{-21pt}\\< \end{array}}
\nc{\gsim}{\begin{array}{c}\sim\vspace{-21pt}\\> \end{array}}
\def\NN{\hbox{\it I\hskip -2.pt N}}
\def\ZZ{\hbox{\it Z\hskip -4.pt Z}}

\def\D{{\cal D}}

\def\hfl#1#2{\smash{\mathop{\hbox to 12mm{\rightarrowfill}}
\limits^{\scriptstyle#1}_{\scriptstyle#2}}}

\def\hfr#1#2{\smash{\mathop{\hbox to 12mm{\leftarrowfill}}
\limits^{\scriptstyle#1}_{\scriptstyle#2}}}

\def\RR{\hbox{\it I\hskip -2.pt R }}
\def\CC{\hbox{\it l\hskip -5.5pt C\/}}

\def\cc#1{\kern .7em\hfill #1 \hfill\kern .7em}

\newcommand{\mysection}[1]{\setcounter{equation}{0}\section{#1}}

\begin{document}
\begin{titlepage}

\begin{center}
April, 1999      \hfill       PM/99-19\\
\vskip .3 in {\large \bf Fractional Supersymmetry~ and}\\
{\large \bf $F^{\mathrm{th}}-$roots of Representations}\\
\vskip .3truecm
{
  {\bf M. Rausch de Traubenberg}\footnote{rausch@lpt1.u-strasbg.fr,
rausch@lpt1.u-strasbg.fr}
   \vskip 0.2 cm
   {\it  Laboratoire de Physique Th\'eorique, Universit\'e Louis Pasteur}\\
   {\it 3-5 rue de l'universit\'e, 67084 Strasbourg Cedex, France}\\ 
  \vskip 0.2 cm
 and \\
   {\it Laboratoire de Physique Math\'ematique et Th\'eorique,
         Universit\'e de Montpellier 2}\\
   {\it place Eug\`ene Bataillon, case 70, 34095 Montpellier Cedex 5, 
     France}\\ 
 \vskip .8 cm
{\bf M. J. Slupinski\footnote{slupins@math.u-strasbg.fr}}\\
{\it Institut de Recherches en Math\'ematique Avanc\'ee}\\
{ \it Universit\'e Louis-Pasteur, and CNRS}\\
{\it 7 rue R. Descartes, 67084 Strasbourg Cedex, France}\\ } 
\end{center}

\vskip .5 in
\begin{abstract}
A generalization of super-Lie algebras
is presented. It is then shown that all known examples of 
fractional supersymmetry
can be understood in this formulation.  However,
the incorporation of three dimensional fractional
supersymmetry in this framework needs some care.
The proposed solutions lead naturally to a formulation of a
fractional supersymmetry starting
from any representation $\D$ of any Lie algebra $g$. This involves taking
the $F^{{\mathrm th}}-$roots of $\D$ in an appropriate sense.  
A fractional supersymmetry in any space-time dimension
is then  possible. This formalism finally 
leads to an infinite dimensional extension
of $g$, reducing to the centerless Virasoro algebra when 
$g=sl(2,\RR)$.
\end{abstract}
\end{titlepage}

\renewcommand{\thepage}{\arabic{page}}

\mysection{Introduction}
Describing   the laws of physics in terms of  underlying symmetries has 
always been a powerful tool. In this respect, it is interesting to study
the kind of symmetries which are allowed in space-time. Within the
framework of Quantum Field Theory (unitarity of the $S$ matrix {\it etc}) 
it is generally admitted that we cannot go beyond supersymmetry (SUSY).
However,
the no-go theorem stating that supersymmetry is {\it the only non-trivial
extension beyond the  Poincar\'e algebra} is valid only if one considers
Lie or Super-Lie algebras. Indeed, if one considers Lie algebras, 
the Coleman and Mandula theorem \cite{cm}  allows only trivial extensions
of the Poincar\'e symmetry, {\it i.e.} extra symmetries must
commute with the Poincar\'e 
generators.
In contrast, if we consider superalgebras, the theorem of 
Haag, Lopuszanski and Sohnius \cite{hls} shows that we can construct a
unique (up to the number of supercharges) superalgebra
extending the Poincar\'e Lie algebra  non-trivially.  It may seem that
these two theorems encompass all possible symmetries of space-time.
But, if one examines the hypotheses of the above theorems, one sees that it
is possible to imagine symmetries which go beyond supersymmetry. Several
possibilities have been considered in the literature \cite{ker, luis,
fsusy, fsusy1d, fr,am, prs, fsusy2d, fvir, fsusy3d, fsusyh}, 
the intuitive idea being
that the generators of the Poincar\'e algebra are obtained as an appropriate
product of more fundamental additional symmetries. These new generators
are in a representation of the Lorentz group which can be neither
bosonic nor fermionic (bosonic charges close under commutators and 
generate a Lie algebra, whilst fermionic charges close under anticommutators
and  induce  super-Lie algebras). 
In this paper we propose an algebraic structure,
called an $F-$Lie algebra, which makes this idea
precise in the context of fractional supersymmetry (FSUSY) of order $F$.
Of course, when $F=1$ this is a Lie algebra, and when 
$F=2$ this is  a Super-Lie algebra.
We show that all examples of FSUSY considered in the literature
can be described within this framework.\\

FSUSY ($F>2$) has been investigated in dimensions one, two and three.
In $1D$ the algebraic structure is relatively simple 
\cite{fsusy1d,fr,am} 
(one just adds a new supercharge $Q$ such that $Q^F=\partial_t$). 
In two dimensions, one can add either two or an infinite number of additional
generators \cite{prs,fsusy2d,fvir}. In three dimensions the situation is  more
complicated. We showed \cite{fsusy3d} that it is possible to inject  
equivariantly the vector representation 
of $so(1,2)$ in a quotient of the $F-$th symmetric product of an appropriate
representation $D_{1/F}$ of $so(1,2)$. In other words, we were able to express
the generators of space-time translations as symmetric $F-$ order polynomials
in more fundamental generators but with the new supercharges
satisfying  extra constraints. 
We also constructed explicitly  in \cite{fsusy3d}
unitary representations of the corresponding algebraic structure
which can be understood as relativistic anyons \cite{lm,b,jn,p}. 
However, it was not possible to consider translations 
as $F-$order symmetric products of the new supercharges without imposing extra
constraints. In contrast,  this problem  exists neither  in dimensions
one and two \cite{fsusy1d,fr,am,prs, fsusy2d}, nor in any dimension when $F=2$
(SUSY). To understand  the results of our paper \cite{fsusy3d}
in terms of $F-$Lie algebras, we propose two solutions: (i) extending the
vectorial representation or, (ii)  extending the 
Poincar\'e algebra  $B= t \oplus so(1,2)$  to $\hat B= \hat t \oplus 
{ \mathrm  Vir}$,
where  $so(1,2) \subset {\mathrm  Vir}$ is the Virasoro 
(without central charge)  
algebra and $\hat t$ a  representation of Vir which extends the
vectorial representation of $so(1,2)$. Correspondingly we also have  to
extend the $D_{1/F}$ representation of $so(1,2)$ to $\hat D_{1/F}$.

The problem encountered in $3D$ FSUSY and especially the solution
we  propose to solve it, enables  us to  define a general method of
associating  an FSUSY to any representation $\D$ of 
any Lie algebra $g$. This algebraic structure 
is in general associated to a non unitary infinite dimensional representation
of $g$. Furthermore, as for $so(1,2)$, one can define
an infinite dimensional Lie algebra $V(g)$ having $g$ as a sub-algebra
and leading to an  $F-$lie algebra.

The content of this paper is as follows. In section two, we give a
precise mathematical definition of the algebraic structure which underlies
FSUSY. Several simple examples are then given. In section 3, we show  
how one can incorporate $3D$ FSUSY into this general mathematical
description by extending the vectorial representation to an appropriate
reducible (but indecomposable) representation.
We then construct FSUSY starting from any
semi-simple Lie algebra $g$ (playing the role of $so(1,2)$) 
and any representation $\D$ (playing the role of the vector representation).
This construction involves taking the $F^{{\mathrm th}}-$root of
$\D$ in some sense.  In particular
this means that one can construct FSUSY in all space-time dimensions.
In section 4, we study an $F-$Lie algebra associated to an infinite dimensional
algebra $V(g)$ having $g$ as a sub-algebra. For $g=so(1,2)$, $V(g)$ reduces
to the centerless Virasoro algebra.

\mysection{Algebraic Structure of Fractional Supersymmetry}

In this section, we give the abstract mathematical structure which
underlies this paper and which generalizes the theory of Lie super-algebras
and their (unitary) representations. Let $F$ be a positive integer and 
$q=\exp{({2i \pi \over F})}$.
We consider a complex vector space $S$ together with a linear map $\varepsilon$
from $S$ into itself satisfying $\varepsilon^F=1$. 
We set $A_k= S_{q^k}$ and $B=S_1$  (where $S_\lambda$  is the eigenspace 
corresponding to the eigenvalue $\lambda$ of $\varepsilon$)
so that $S=B\oplus_{k=1}^{F-1} A_k$. The map $\varepsilon$ is called the 
grading.
If $S$ is endowed with the following structures we will say that $S$ is
a fractional super Lie algebra ($F$-Lie algebra for short):

\begin{enumerate}
\item $B$ is a Lie algebra and $A_k$ is a   representation of $B$.
We write these  structures as a bracket $[b,X]$ with the understanding that
$[b,X]=-[X,b]$ if $X \in A_k,~ b\in B$. It is clear that 
$[\varepsilon(X),\varepsilon(Y)]= \varepsilon\left([X,Y]\right)$.  
\item There are multilinear,  $B-$equivariant  ({\it i.e.} which respect 
the action  of  $B$) maps
 $\left\{~~, \cdots,~~ \right\}: {\cal S}^F\left(A_k\right) 
\rightarrow B$ from 
${\cal S}^F\left(A_k\right)$ into  $B$.
In other words, we  assume that some of the elements of the Lie algebra $B$ can
be expressed as $F-$th order symmetric products of
``more fundamental generators''. Here  ${ \cal S}^F(D)$ denotes 
the $F-$fold symmetric product
of $D$. It is then easy to see that: 

\beqa
\label{eq:epsi}
\left\{\varepsilon(a_1), \cdots, \varepsilon(a_F)\right\}=
\varepsilon\left(\left\{a_1, \cdots, a_F\right\}\right), 
\forall a_i \in A_k.
\eeqa

\item  For $b_i \in B$ and $a_j \in A_k$ the following 
``Jacobi identities'' hold:

\beqa
\label{eq:J}
&&\left[\left[b_1,b_2\right],b_3\right] + 
\left[\left[b_2,b_3\right],b_1\right] +
\left[\left[b_3,b_1\right],b_2\right] =0 \nonumber \\
&&\left[\left[b_1,b_2\right],a_3\right] +
\left[\left[b_2,a_3\right],b_1\right] +
\left[\left[a_3,b_1\right],b_2\right]  =0 \nonumber \\
&&\left[b,\left\{a_1,\dots,a_F\right\}\right] =
\left\{\left[b,a_1 \right],\dots,a_F\right\}  +
\dots +
\left\{a_1,\dots,\left[b,a_F\right] \right\} \\
&&\sum\limits_{i=1}^{F+1} \left[ a_i,\left\{a_1,\dots,
a_{i-1},
a_{i+1},\dots,a_{F+1}\right\} \right] =0. \nonumber
\eeqa

\noindent
The first identity is the usual Jacobi identity for Lie algebras,
the second says that the $A_k$ are representation spaces of $B$ and
the third is just the Leibniz rule (or the equivariance of  
 $\left\{~~, \cdots,~~ \right\}$). The fourth identity is
the analogue of the graded Leibniz rule of Super-Lie algebras
for $F-$Lie algebras

\hskip -1 cm 
If we want to be able to talk about unitarity, we also require the
following additional 
struc-

\hskip -1 cm
ture and in this case, $S$ is called an
$F-$Lie algebra with adjoint.

\item
A conjugate linear map $\dag$ from $S$ into itself  such that:

\beqa
\label{eq:conj}
\begin{array}{ll}
\mathrm{a)}&(s^\dag)^\dag=s, \forall s \in S \cr
\mathrm{b)}& \left[a,b\right]^\dag= \left[b^\dag,a^\dag\right] \cr 
\mathrm{c)}&\varepsilon(s^\dag)=\varepsilon(s)^\dag \cr
\mathrm{d)}& \left\{a_1,\cdots,a_F\right\}^\dag=
 \left\{\left(a_1\right)^\dag,\cdots,\left(a_F\right)^\dag\right\},~~~
\forall a \in A_k.
\end{array}
\eeqa

From a) and c) we see that for 
$X \in B$ we have  $X^\dag \in B$, and that for
  $X \in A_k$, we have   $ X^\dag \in  A_{F-k}.$
\end{enumerate} 
\noindent

A unitary representation of an $F-$Lie algebra
with adjoint $S$  is  a linear map
$\rho : ~ S \to \mathrm{End}(H)$, 
(where $H$ is a Hilbert space
and ${\mathrm{End}}(H)$ the space of linear operators acting on $H$) 
and a unitary
endomorphism $\hat \varepsilon$ such that $ \hat \varepsilon^F=1$ 
which satisfy 

\beqa      
\label{eq:rep}
\begin{array}{ll}
\mathrm{a)}& \rho\left(\left[x,y\right]\right)= \rho(x) \rho(y)- 
\rho(y)\rho(x) \cr
\mathrm{b)}& \rho \left\{a_1.\cdots,a_F\right\}=
{1 \over F !} \sum \limits_{\sigma \in S_F}
\rho\left(a_{\sigma(1)}\right) \cdots \rho\left(a_{\sigma(F)}\right) \cr
\mathrm{c)}& \rho(s)^\dag = \rho(s^\dag) \cr
\mathrm{d)}& \hat \varepsilon \rho\left(s\right) \hat \varepsilon^{-1} =
\rho\left(\varepsilon\left(s\right)\right)
\end{array}
\eeqa

\noindent
($S_F$ being the group of permutations of $F$ elements). Note that
with the normalisation of b), when $F=2$, one has 
$\rho(\{a_1,a_2\})=1/2(a_1 a_2 + a_2 a_1)$ instead of the usual
$\rho(\{a_1,a_2\})=(a_1 a_2 + a_2 a_1)$. 
As  a consequence of these properties, 
since the eigenvalues of $\hat \varepsilon$ are $\mathrm{F}^{\mathrm{th}}-$
roots of unity, we have  the following decomposition of the Hilbert space

$$H= \bigoplus \limits_{k=0}^{F-1} H_k,$$

\noindent
where $H_k=\left\{\left|h\right> \in H ~:~ 
\hat \varepsilon\left|h\right>=q^k \left|h\right> \right\}$.
The operator $N \in \mathrm{End}(H)$
(the set of linear operators acting on $H$) 
defined by $N\left|h\right>=k
\left| h \right>$ if $\left|h\right> \in H_k$
is the  ``number operator'' (obviously $q^N=\hat \varepsilon$).
Since $\hat \varepsilon \rho(b)= \rho(b) \hat \varepsilon, \forall b \in B$
each $H_k$ provides a representation of the Lie algebra $B$. 
Furthermore, for $a \in A_\ell$,
 $\hat \varepsilon \rho(a)=q^\ell \rho(a) \hat \varepsilon$ and so
we have 
$\rho(a) .H_k\ \subseteq 
H_{k+\ell ({\mathrm{mod~} F)}}$  \\

\noindent
\underline{Remark 1}:

\noindent
For all $k=1,\cdots, F-1$ it is clear that 
the subspace  $B \oplus A_k$ of $S$ satisfies (\ref{eq:epsi}-\ref{eq:J}) 
and the subspace $B \oplus A_k \oplus A_{-k}$ satisfies  
(\ref{eq:epsi}-\ref{eq:conj}) (when $S$ has an adjoint).

\noindent
\underline{Remark 2}:

\noindent
It is important to notice
that bracket $\{ \cdots \}$ is a priori  not defined for  elements in different
gradings. 

\noindent
\underline{Remark 3}:

\noindent 
If we set 

\beqa
\label{eq:real}
 B_{{\tiny \RR}}  =\Big \{b \in B: b^\dag=-b \Big \} \nonumber \\
 A_{{\tiny \RR}}  =\Big \{a \in A: a^\dag=a \Big \}, 
\eeqa

\noindent
then $S_{{\tiny \RR}} = B_{{\tiny \RR}} \oplus A_{{\tiny \RR}}$ is stable by 
$\varepsilon$ and satisfies  (\ref{eq:epsi}-\ref{eq:J}). 
Here we use the normalizations conventionally used in mathematical literature
(no $i$ factor in the structure constants of the algebra). For physicists,
notice that if $b^\dag = -b$ then $(i b)$ is hermitian.

\vskip .5truecm
 
\subsection*{Example 1:}

Obviously, a $1-$Lie algebra is just a Lie algebra.
A $2-$Lie algebra is just  a  Lie super-algebra:
$S=B\oplus A_{1}$, with  even part $B$  and 
odd part $A_1$. In supersymmetry,
because of the spin statistics theorem $A_1$ is a fermionic 
representation of $B$. Note that the  Jacobi identities
(\ref{eq:J}) above reduce to the standard Jacobi identities of a Super-Lie
algebra. If we consider unitarity in SUSY,
property (\ref{eq:conj}), (\ref{eq:rep}) above have also to be considered but
for the super-Poincar\'e algebra,
the nature of $\dag$ depends very much on the dimension 
of the space-time and the signature of 
the metric (Majorana, Weyl, Majorana-Weyl, $SU(2)-$Majorana and
$SU(2)-$Majorana-Weyl conditions).

\vskip .5truecm
\subsection*{Example 2:}

Let $V$ be finite dimensional complex vector space and let
$\varepsilon:~V \rightarrow V$ be a linear operator satisfying
$\varepsilon^F=1$. Then
 
$$V=\oplus_{k=0}^{F-1} V_k,$$

\noindent
where $V_k= \left\{ \left|v\right> \in V : \varepsilon \left|v\right>=q^k
\left|v\right> \right\}$. We define

$$A_k=\left\{ f \in \mbox{End}(V) : \varepsilon\circ f\circ \varepsilon^{-1}=
 q^k f\right\}$$

\noindent
and  $S=B \oplus_{k=1}^{F-1} A_k$ (with $B=A_0$). Since 
$A_k A_\ell \subset A_{k + \ell (\mbox{\tiny{mod~}} F)}$ one has

\beqa
\label{eq:Vect1}
&\left[A_0, A_k \right] \subset A_k \\
\label{eq:Vect2}
&{\underbrace{A_k A_k \cdots A_k}} \subset A_0. \\
& \hskip -1cm {\mathrm F-times} \nonumber
\eeqa   

\noindent
The bracket $[~,~]$  of $S$ is defined by  (\ref{eq:Vect1}) and  
$\{ \cdots \}: S^F(A_k) \rightarrow B$ by \\
$\left\{a_1 \cdots a_F\right\} =
1/F! \sum \limits_{\sigma \in S_F} a_{\sigma(1)} \cdots a_{\sigma(F)}$.
The first three Jacobi (\ref{eq:J}) identities are clearly
satisfied, and  calculation shows that
the last Jacobi identity also holds.
Thus $S$ is an $F-$Lie algebra. If $V$ is endowed with 
a hermitian metric and $\varepsilon$ is a unitary operator
then adjunction defines an adjoint on the $F-$Lie algebra $S$. 

\vskip .5truecm
\subsection*{Example 3:}

In $1D$ \cite{fsusy1d,fr,am} the simplest 
$F-$Lie algebra is two dimensional,
and is generated by the operators $\partial_t, Q$ with the  relation 
$Q^F=\partial_t$. We take $B= \Big < \partial_t \Big >$,
the translation in time, and $A_1 = \Big < Q \Big >$ . We obviously have
$\varepsilon (\partial_t)=\partial_t$ and $\varepsilon(Q) = q Q$. 
An explicit representation in terms of  generalized grassmann variables
\cite{rr,r,re,frr,hq} 
can be constructed \cite{fsusy1d,fr,am}.
It is possible to extend this $F-$Lie algebra to a $F-$Lie 
algebra with adjoint
by the addition of one more 
generator, $Q^\dag$, such that $\varepsilon(Q^\dag)=q^{-1} Q^\dag$,
$(Q^\dag)^F=(\partial_t)^\dag=-\partial_t$. 
Let us recall once again that there are no 
algebraic relations between $Q$ and $Q^\dag$.

\vskip .5truecm
\subsection*{Example 4:}

In $2D$ there are  several possible algebras. The simplest one
is obtained by considering the $3$ generators $\partial_z, \partial_{\bar z}$
and $Q_z$. We set 
 $B= \Big < \partial_z, \partial_{\bar z}  \Big >$ and 
$A_1= \Big <Q_z \Big >$, the relations are $ \big ( Q_z \big)^F= \partial_z$
 and $[\partial_z, Q_z]= [\partial_{\bar z}, Q_z]=0$. \\
This $F-$Lie algebra can be extended to a
$4-$dimensional $F-$Lie algebra with adjoint 
by adding one more generator $Q_{\bar z} \in A_{-1}$ such that
$\left(\partial_z\right)^\dag=\partial_{\bar z}$,
$\big ( Q_{\bar z} \big)^F=\partial_{\bar z}$ and 
$\big ( Q_z \big)^\dag= Q_{\bar z}$ 
\cite{prs,fsusy2d}. \\
There is also a more complicated algebraic extension, involving an infinite
number of generators which corresponds to an extension of the Virasoro
algebra without central charge. In addition to the Virasoro generators $L_n,
\overline{L}_n~
n \in \ZZ$ we add the generators $G_r, r \in \ZZ+1/F$, which
correspond to the modes of  a field of conformal weight $1+1/F$ and satisfy the
following relations \cite{fvir}   

\beqa
\label{eq:fv2}
\left[L_n,L_m\right]& =& (n-m)L_{m+n} \nonumber \\
\left[\overline{L}_n,\overline{L}_m\right]& =& (n-m)\overline{L}_{m+n} 
\nonumber \\
\left[{L_n},\overline{L}_m\right]&=& 0 \\
\left[L_n,G_r\right] &=& ({n \over F}-r)G_{n+r} \nonumber \\
\left[\overline{L}_n,G_r\right] &=&0 \nonumber \\
\left\{G_{r_1},\cdots,G_{r_F}\right\}&=& L_{r_1 + \cdots + r_F}, \nonumber 
\eeqa

\noindent
Here we take $B= {\mathrm {Vir}} \oplus \overline{{\mathrm {Vir}}}$
and $A_1=\Big < G_r, r \in \ZZ + 1/F \Big >$. In this extension, we
have $L_1 \equiv \partial_z, \overline{L}_1 \equiv \partial_{\bar z}$
and $G_{{1 \over F}} \equiv  Q_z$. We can also include an adjoint by
adding $A_{-1}=\overline{A}_1$. As in $1D$, it
is possible to construct an explicit realisation of the above algebras 
using  generalised grassmann  variables \cite{fvir,prs, fsusy2d}.  

In all the given examples, appropriate representations have been obtained in
terms of adapted superfields \cite{fsusy1d,fr,am,prs,fsusy2d}. To our
knowledge, unitarity remains an unsolved problem.

\vskip .5truecm
In three dimensions, the situation is much more complicated and we will
study this in the next section.

\subsection*{Example 5}
Let $g$ be a complex Lie algebra and let ${\mathbf r}, {\mathbf r^\prime}$
be representations of $g$ such that there is a $g-$equivariant map
$\mu: S^F(\mathbf{r}) \rightarrow \mathbf{r^\prime}$.  We set
 
$$S=B\oplus A_1 = (g \oplus {\mathbf r^\prime}) \oplus    {\mathbf r}.$$

\noindent
$B=g \oplus {\mathbf r^\prime}$ is a  Lie algebra 
as the semi-direct product of $g$ and $\mathbf{r^\prime}$ 
(the latter with the trivial bracket). We can extend the action of  $g$ on
${\mathbf r}$  to an action of $B$ on  ${\mathbf r}$
by letting ${\mathbf r^\prime}$ act trivially on ${\mathbf r}$.
This defines the bracket $[~,~]$ on $S$. For the map $\left\{\cdots \right\}$
we take $\mu$. The first three Jacobi identities (\ref{eq:J})
are clearly satisfied, and the fourth is also satisfied as
each term in the expression of the L.H.S  vanishes.

For example, if    

$${\cal S}^F\left({\mathbf r}\right)= \oplus_k {\mathbf r_k},$$

\noindent
is a decomposition into irreducible summands, then for a given $k$

$$S_k=(g \oplus {\mathbf r_k}) \oplus {\mathbf r},$$
is an  $F-$Lie algebra. \\ 
As an illustration, if $g=so(1,2)$ and
${\mathbf r}={\mathbf 2} \oplus {\mathbf 1} $ (the spin representation plus
the trivial representation), then 
${\cal S}^3({\mathbf 2} \oplus {\mathbf 1})
={\cal S}^3({\mathbf 2}) \oplus  {\cal S}^2({\mathbf 2}) \oplus
{\mathbf 2} \oplus {\mathbf 1} = {\mathbf 4} \oplus 
 {\mathbf 3} \oplus  {\mathbf 2} \oplus  {\mathbf 1} $ and it is possible to 
obtain the spinorial or the vectorial representations  of $so(1,2)$ 
from a symmetric product of order 3. This can be compared with 
the result of R. Kerner \cite{ker} where a cubic root of the Dirac equation
is obtained. More generally, for any $F$:

${\cal S}^F({\mathbf 2} \oplus {\mathbf 1})={\cal S}^F({\mathbf 2}) \oplus
{\cal S}^{F-1} ({\mathbf 2}) \oplus \cdots \oplus {\mathbf 1} =
(\mathbf{F+1}) \oplus \mathbf{F} \oplus \cdots \oplus \mathbf{1}$. 

\mysection{Fractional Supersymmetry and finite dimensional \ Lie algebras}
\subsection{Fractional Supersymmetry in three dimensions}

In \cite{fsusy3d} we considered FSUSY in three dimensions. 
In order to understand our results in term of $F-$Lie algebras
let us introduce a realisation of $so(1,2)$ which is convenient 
for explicit calculations. Let ${\cal F}$ be the vector space
of functions on $\RR^{2,*}_+=\big\{(x,y) \in \RR^2: x,y > 0 \big \}$.
Consider the linear operators acting on ${\cal F}$ given by

\beqa
\label{eq:so(1,2)}
J_{-} &=& x \partial_y \nonumber \\
J_0 &=& {1 \over 2} (y \partial_y - x \partial_x) \\
J_{+} &=& y \partial_x. \nonumber 
\eeqa

\noindent
These operators satisfy the commutation relations

\beqa
\label{eq:corel}
\big[J_{-}, J_+\big] &=& -2 J_0 \nonumber \\
\big [J_{0}, J_+\big] &=& J_+  \\
\big[J_{0}, J_{-}\big] &=& -J_{-},  \nonumber
\eeqa

\noindent
and thus generate the Lie algebra $so(1,2)$.
It is easy to check that the following  subspaces of ${\cal F}$ are 
representations of $so(1,2)$:

\beqa
\label{eq:repp}
\D_{-n}&=&\Big<~ x^{2n},x^{2n-1}y,\cdots, x y^{2n-1}, y^{2n} \Big >,
\hskip .5cm \Big(n \in \NN/2 \Big)
\nonumber \\
\D^+_{-\lambda}&=&\Big<~ x^{2 \lambda}  \left( {y \over x }\right)^m, 
m \in \NN \Big >,\hskip .5cm 
\Big(\lambda \in \RR \setminus \NN/2\Big) \\
\D^-_{-\lambda}&=&\Big<~ y^{2 \lambda} 
\left( {x \over y }\right)^m, m \in \NN \Big >, \hskip .5cm
\Big( \lambda \in \RR \setminus \NN/2 \Big). \nonumber   
\eeqa

\noindent
Of course other representations can also be obtained (for instance unbounded
from below and above) but they are not useful for our purpose.

The  representation $\D_{-n}$ is the $(2n+1)-$dimensional  irreducible 
representation and the representations $\D^\pm_{-\lambda}$ are infinite
dimensional representations, bounded from below and above respectively.
It is important to emphasize that the representations given in (\ref{eq:repp})
do not have the normalizations  conventionally used in the literature and 
the basis is not orthonormal, but those normalizations are convenient
for further developments.
For a general classification of the representations
of  three-parameter Lie algebras,  see {\it e.g.} \cite{wy} where analogous
monomials (of the form
$x^\alpha y^\beta \left({x \over y}\right)^m$, with $\alpha, 
\beta \in \CC, m \in \ZZ)$ are considered.  \\
In the paper \cite{fsusy3d} we introduced four representations, 
$\D^\pm_{-1/F,\pm}$. These are are related to the above representations
by the following isomorphisms 

\beqa
D^+_{-1/F,+} \cong D^+_{-1/F,-} \cong D^+_{-1/F} \nonumber \\
D^-_{-1/F,+} \cong D^-_{-1/F,-} \cong D^-_{-1/F}. \nonumber
\eeqa

\noindent
In this article, for practical reasons we work only with the representations
(\ref{eq:repp}).

The multiplication map 
$m_n : {\cal F} \times \cdots \times {\cal F} \rightarrow 
{\cal F}$ given by 

\beq
\label{eq:multi}
 \ m_n(f_1,\cdots, f_n)= f_1 \cdots f_n
\eeq

\noindent
 is multilinear and 
totally symmetric and hence induces a map $\mu_F$ from
${\cal S}^F({\cal F})$ into ${\cal F}$. Restricting to 
${\cal S}^F\left(D^\pm_{-1/F}\right)$ one sees that

\beqa
\label{eq:FSUSY}
S^F\big(\D^{+}_{-1/F}\big)_{{\mathrm red}} \buildrel{\hbox{def}} \over = 
\mu_F\Big(S^F\big( D^{+}_{-1/F}\big)\Big) 
&=&\Big<x^2 
\left({y \over x} \right)^m, ~ m\in \NN ~~\Big> \supset \D_{-1} \\
S^F\big(\D^{-}_{-1/F}\big)_{{\mathrm red}} \buildrel{\hbox{def}} \over = 
\mu_F\Big(S^F\big( D^{-}_{-1/F}\big)\Big) 
&=&\Big<y^2 
\left({x \over y} \right)^m, ~ m\in \NN ~~\Big> \supset \D_{-1} \nonumber
\eeqa  

\noindent
The simple observation  of (\ref{eq:FSUSY}) together with example 5 
in section 2 naturally lead to the $F-$Lie algebra

\beqa
\label{eq:F-so(1,2)}
\Big(so(1,2) \oplus S^F\big(\D^{\pm}_{-1/F}\big)_{{\mathrm red}} \Big)
\oplus \D^\pm_{-1/F}.
\eeqa

\noindent
In \cite{fsusy3d}, by considering an adapted conjugations $\dag:
\D^+_{-1/F}= \left(\D^-_{-1/F}\right)^\dag$ and from the Wigner induced
representation we proved that the representations of 3d-FSUSY are unitary
and induce a symmetry between relativistic anyons.

Looking at the representations defined in (\ref{eq:FSUSY}) {\it i.e.}
$S^F\big(\D^{\pm}_{-1/F}\big)_{{\mathrm red}}$,
one sees that,  even though $\D_{-1}$ is a subspace stable under
 $so(1,2)$ there is no  complement stable under $so(1,2)$
\cite{fsusy3d}. Indeed,
these representations cannot be built from a primitive vector.
This is due to the fact that $J_{+}^3 \Big(x^2\Big) =0$ 
and consequently we cannot reach $x^{-1}y^3$ from $x^2$
but conversely $J_{-}^3 $ $ \Big(x^{-1}y^3)= 6 x^2$ (such reducible but 
indecomposable
representations also appear in \cite{wy}). 
This is the reson why there is no $F-$Lie algebra structure on
$so(1,2) \oplus \D_{-1}$.

\subsection{Extension to any Lie algebra}
We consider now $g$ a complex semi-simple Lie algebra 
of rank $r$ and $\D$ an arbitrary 
representation. The purpose of this section, is to construct an $F-$Lie
algebra $S=B \oplus A_1$ such that the Lie algebra $B$ contains
the semi-direct product $g \oplus \D$. If $g=so(1,2)$ and $\D$ is
the vector representation, this construction leads to the $F-$Lie algebra
(\ref{eq:F-so(1,2)}) above.

Let
$h$ be a Cartan sub-algebra of $g$, let $\Phi \subset h^\star$ (the dual of
$h$)  be the corresponding set of roots and let
$f_\alpha$ be the one dimensional root space associated to $\alpha \in \Phi$.
We choose a basis  $\{H_i, i=1, \cdots,  r\}$ of $h$ and elements 
$E^\alpha \in f_\alpha$ such that the commutation relations  become

\beqa
\label{eq:lie}
\big[H_i,H_j \big] &=& 0 \nonumber \\
\big[H_i,E^\alpha \big] &=&  \alpha^i E^\alpha \\
\big[E^\alpha, E^\beta\big] &=& \left \{
\begin{array}{ll}
\epsilon\{\alpha,\beta\} E^{\alpha+\beta}& {\mathrm {~~if~~}} 
\alpha + \beta \in \Phi \cr
{2\alpha.H \over \alpha.\alpha}& {\mathrm {~~if~~} } \alpha+\beta=0 \cr
0& {\mathrm {~~otherwise }}
\end{array}
\right. \nonumber
\eeqa

\noindent
Recall that the real Lie algebra spanned by the $H_i$ and the
$E^\alpha$ is the split real form of $g$, and that the real Lie
algebra spanned by $iH_j$, $E^\alpha-E^{-\alpha}$ and
$i(E^\alpha+E^{-\alpha})$ is the  compact real form of $g$.

We now introduce $\{\alpha_{(1)},\cdots, \alpha_{(r)}\}$ (the positive roots)
a basis of simple roots.  
The weight lattice $\Lambda_W(g) \subset h^\star$ is the set of vectors
$\mu$ such that $ {2 \alpha.\mu \over \alpha.\alpha} \in \ZZ$ and, as is
well known, there is a  basis of the weight lattice consisting of the
fundamental weights $\{\mu_{(1)}, \cdots, \mu_{(r)} \}$ defined by
$ {2 \mu_{(i)}.\alpha_{(j)} \over \alpha_{(j)}.\alpha_{(j)}}= \delta_{ij}$.
A weight $\mu = \sum \limits_{i=1}^r n_i \mu_{(i)}$ is called dominant if
all the $n_i \ge 0$ and it is well known that the set of dominant weights
is in one to one correspondence with the set of (equivalence classes of)
irreducible finite dimensional representations of $g$.

Recall briefly how one can associate a  representation of $g$ to 
$\mu \in h^\star$. 
In ${\cal{U}}(g)$, the universal enveloping
algebra of $g$, let $I_\mu$ be the left-ideal  generated by the elements 

$\Big\{E^\alpha  (\alpha > 0), h_i-\mu(h_i). {\mathbf I} (h_i \in h)\Big\}$, 
where 
$ {\mathbf I}$ is the identity of ${\cal{U}}(g)$ and $h_i= 2 
{\alpha_{(i)}.H_i \over \alpha_{(i)}^2}$.

\noindent
The Lie algebra $g$ acts on ${\cal{U}}(g)$ by left multiplication,  $I_\mu$ is
stable under this action and therefore the quotient 
${\cal V}_\mu={\cal{U}}(g)/I_\mu$ is a representation space of $g$:
${\cal V}_\mu$ is a highest weight representation and is called
the Verma module associated to $\mu$ \cite{kr}. If $\mu$ is
dominant, then ${\cal V}_\mu$ has a unique maximal proper 
sub-representation $M_\mu$ and the quotient $\D_\mu= {\cal V}_\mu/M_\mu$
is an irreducible finite dimensional representation of $g$.

To come back to our original  problem, consider a finite dimensional 
irreducible
representation $\D_\mu$ of $g$. If ${\cal V}_{\mu/F}$ is the Verma
module associated to $\mu/F$, there is a $g-$equivariant map

\beq
\label{eq:sfmu}
i: {\cal V}_\mu \rightarrow {\cal S}^F\left({\cal V}_{\mu/F} \right),
\eeq

\noindent
because ${\mathbf I} \otimes \cdots \otimes {\mathbf I} \in 
{\cal S}^F\left({\cal V}_{\mu/F}\right)$ is a highest weight vector of 
weight $\mu$.
Taking the quotient by $M_\mu$ one obtains a $g-$equivariant
inclusion 

\beq
\label{eq:dmu}
\D_\mu \hookrightarrow {\cal S}^F\left({\cal V}_{\mu/F} \right)/i(M_\mu),
\eeq
  
\noindent
since $D_\mu= {\cal V}_\mu/M_\mu$. Denoting 
$ {\cal S}^F\left({\cal V}_{\mu/F} \right)/i(M_\mu)$ by
${\cal S}^F\left(\D_{\mu/F}\right)_{\mathrm{red}}$ then, as in example 5,

\beq
\label{eq:F-g}
S=(g \oplus {\cal S}^F\left(\D_{\mu/F}\right)_{\mathrm{red}}) \oplus
{\cal V}_{\mu/F}
\eeq

\noindent
is naturally an $F-$Lie algebra.

We can reformulate this construction in less abstract terms.
If $\left|\mu\right>$ is the primitive vector
associated to a dominant weight $\mu$ ({\it i.e.}
($E^\alpha \left|\mu\right> =0, \alpha >0$ and $h_i\left|\mu\right>=
n_i\left|\mu\right>=
\mu(h_i)\left|\mu\right>$)),  the representation 
$\D_\mu$ is generated by the action of the $E^\alpha, \alpha < 0$ on 
$\left|\mu\right>$.  
Because the representation is finite dimensional, corresponding 
to the highest weight state $\left|\mu\right>$ we have a
lowest state $\left|\mu^\prime\right>$, 
$\mu^\prime=  \sum \limits_{i=1}^r n^\prime_i \mu_{(i)}$
 ($E^\alpha\left|\mu^\prime\right>=0, \alpha <0$). 
Of course the two representations built with $\mu$ or $\mu^\prime$
are the same. But, if the weight is not dominant the situation is
more involved. For our purposes, to the representation $\cal{D}_\mu$
we associate two infinite dimensional representations: one  associated
to the weight ${\mu \over F}$ and one to ${\mu^\prime \over F}$, noted
${\cal{D}}_{{\mu \over F}}^\pm$ respectively
(the first is bounded from below and the second from above). 
These two inequivalent  representations are characterized by primitive vectors:

\beqa
\label{eq:muF}
{\cal{D}}_{{\mu \over F}}^+: \left|{\mu \over F} \right>&&
h_i\left|{\mu \over F} \right> = {n_i \over F} \left|{\mu \over F} \right>, 
E^\alpha \left|{\mu \over F} \right> =0, \alpha >0 \\
{\cal{D}}_{{\mu \over F}}^-: \left| {\mu^\prime \over F} \right>&&
h_i\left|{\mu^\prime \over F} \right> = {n^\prime_i \over F} 
\left|{\mu^\prime \over F}\right>, 
E^\alpha \left|{\mu^\prime \over F} \right> =0, \alpha < 0. \nonumber
\eeqa
 
\noindent
$\D^+_{\mu/F}$ is the Verma module ${\cal V}_{\mu/F}$  
abstractly defined above:
In these terms the projections $\mu_F$ from  
${\cal S}^F\Big( \D^\pm_{\mu \over F}\Big)  \longrightarrow
{\cal{S}}^F\left({\cal{D}}_{\mu \over F}^\pm\right)_{{\mathrm red}}$ is given
by $\mu_F \left({\cal S}^F\Big(\left|h_1\right>,\cdots,\left|h_F\right>
\Big) \right)= \left|h_1 + \cdots + \cdots h_F \right>$.

\noindent
We observe that that ${\cal{D}}_\mu \subset
{\cal{S}}^F\left({\cal{D}}_{\mu \over F}^\pm\right)_{{\mathrm red}}$
because $\mu_F \left({\cal S}^F\Big(\left|{\mu \over F}\right>,\cdots,
\left|{\mu \over F}\right> \Big) \right)= \left|\mu\right>$ 
(or $\mu_F \left( {\cal S}^F\right. $ $\left. 
\Big( \left| {\mu^\prime  \over F} \right>, 
\cdots,
\left|{\mu^\prime \over F}\right> \Big) \right)= \left|\mu^\prime\right>$)
is the primitive vector of $\D_\mu$.
  
\section{Fractional supersymmetry and infinite dimensional algebras}
In the previous section we  constructed a canonical
$F-$Lie algebra 

\beq
\label{eq:flie}
S=(g \oplus {\cal S}^F\left(\D_{\mu/F}\right)_{\mathrm{red}}) \oplus \D_{\mu/F}
\eeq

\noindent
associated to a finite dimensional Lie algebra $g$ 
and an irreducible finite dimensional representation $\D_\mu$. 
In this section we will
show that one can extend the representation  $g$ in $D_\mu$  to a 
representation  of an infinite dimensional Lie algebra 
$V(g)$ in $\hat D_\mu$, and construct
an $F-$Lie algebra containing (\ref{eq:flie}) as a sub-algebra.

\subsection{$so(1,2)$ and the Virasoro algebra}

It is well know that the Virasoro algebra admits $so(1,2)$ as a sub-algebra.
The action  (\ref{eq:so(1,2)}) of $so(1,2)$ on ${\cal F}$ extends to
an action of the Virasoro algebra (without central extension) on
${\cal F}$ by setting

\beqa
\label{eq:vir}
L_n={n+1 \over 2}  \left( {y \over x} \right)^n x \partial_x
+{n-1 \over 2}  \left( {y \over x} \right)^n y \partial_y, n\in \ZZ.
\eeqa

\noindent
One can verify the commutation relations

\beqa
\label{eq:comrel2}
[L_n,L_m]=(n-m) L_{n+m}
\eeqa

\noindent
and that $J_{-}=-L_{-1},J_0=-L_0,J_+=L_1$.

In analogy with  the representations (\ref{eq:repp}) of $so(1,2)$ we
define representations of Vir as follows:

\beqa 
\label{eq:repv}
\hat\D_{-n}&=&\Big<~ f^{(-n)}_{m-n}=x^{2n} \left( {y \over x} \right)^m, m 
\in \ZZ   \Big >, \hskip .5cm \left(n \in \NN/2 \right) \nonumber \\
\hat \D^+_{-\lambda}&=&\Big<~ f^{(+,-\lambda)}_{m-\lambda}=x^{2 \lambda} 
\left( {y \over x }\right)^m, m \in \ZZ \Big >,
\hskip .5cm \left(\lambda \in \RR \setminus \NN/2 \right)\\
\hat \D^-_{-\lambda}&=&\Big<~ f^{(-,-\lambda)}_{\lambda -m}=y^{2 \lambda} 
\left( {x \over y} \right)^m, m \in \ZZ \Big >,
\hskip .5 cm \left(\lambda \in \RR \setminus  \NN/2 \right). \nonumber
\eeqa
 
\noindent
Then, one can check explicitly that the action of Vir on (\ref{eq:repv})
is given by

\beqa
\label{eq:viract}
L_k \left(f^{(-n)}_{p}\right)&=&(kn-p) f^{(-n)}_{k+p} \nonumber \\
L_k \left(f^{(+,-\lambda)}_{p}\right)&=&(k\lambda-p) 
f^{(+,-\lambda)}_{k+p} \\
L_k \left(f^{(-,-\lambda)}_{p}\right)&=&(k\lambda-p)  
f^{(+,-\lambda)}_{k+p}, \nonumber
\eeqa

\noindent 
where the indices in (\ref{eq:repv}) are chosen in such a way that  they 
correspond
to the eigenvalues of $-L_0$  {\it i.e.} the helicity. In the language of
conformal field theory, $f^{(-n)}_p,  f^{(+,-\lambda)}_p$ and 
$f^{(-,-\lambda)}_p$ correspond to the modes of  conformal fields of 
conformal weight
$n+1$ and $\lambda+1$ respectively.

Finally, we observe  that the the representations (\ref{eq:repp}) are included
in the corresponding representations (\ref{eq:repv}) and  in each case
that the action
of Vir extends the action of $so(1,2)$. Let us remark that these 
representations   are all unbounded from below and above
({\it i.e.} they cannot be obtained from primitive vectors or 
a highest/lowest weight
state).\\

The fundamental property of these representations is that there is a
Vir-equivariant map from $S^F\Big(\hat \D^+_{-1/F}\Big)$ and 
$S^F\Big(\hat \D^-_{-1/F}\Big)$
to $\hat \D_{-1}$. This is just the multiplication map
$\mu_F: {\cal F} \rightarrow {\cal F}$ 
(see (\ref{eq:multi})) which is obviously  Vir-equivariant.
In fact,  a direct calculation shows that

\beqa
\label{eq:FSUSY1}
S^F\big(\hat\D^{\pm}_{-1/F}\big)_{{\mathrm red}} \buildrel{\hbox{def}} \over = 
\mu_F\Big(S^F\big(\hat\D^{\pm}_{-1/F}\big)\Big) 
 \cong \hat \D_{-1},
\eeqa  

\noindent
and hence that 
$S^F\big(\hat\D^{\pm}_{-1/F}\big)_{{\mathrm red}}$ and $\hat \D_{-1}$
are isomorphic.

By the method explained in example 5, 

\beq
\label{eq:falg}
S=\left(\mathrm{Vir} \oplus \hat \D_{-1}\right) \oplus 
\hat D^{+}_{-1/F} \oplus \hat D^{-}_{-1/F},
\eeq

\noindent
is an $F-$Lie algebra.





\noindent
Denoting $\hat \D_{-1} = \Big<~P_{m-1}=f^{(-1)}_{m-1}, m\in \ZZ ~~\Big>$, 
$\hat \D^\pm_{-1/F} = \Big<~Q^\pm_{\pm(m-1/F)}=f^{(\pm,-1/F)}_{\pm(m-1/F)}, 
m\in \ZZ ~~\Big>$, the brackets in $S$ are given explicitly 
by the following formulae:

\beqa
\label{eq:F-vir}
\Big[ L_n, L_m \Big] &=& (n-m) L_{n+m} \nonumber \\
\Big[ L_n, P_m \Big] &=&(n-m) P_{n+m} \nonumber \\
\Big[ L_n, Q^\pm_r \Big] &=&({n  \over F}-r) Q^\pm_{n+r}\nonumber \\
\Big[ P_n, P_m \Big]  &=& 0 \\
\Big[ P_n, Q^\pm_r \Big]  &=& 0 \nonumber \\
\left\{Q^\pm_{r_1}, \cdots,Q^\pm_{r_F} \right\}  \nonumber
&=&P_{r_1+ \cdots r_F}. 
\eeqa






\noindent
\underline{Remark}
Any weight $\mu$ of the Lie algebra $so(1,2)$ can be considered as  
a weight 
$\hat \mu$ of Vir if we set $\hat \mu(L_0)= -\mu(J_0)$ \cite{kr}.
As in section 3.1, we can construct the associated Verma module
and define the $F-$Lie algebra

\beq
\label{eq:F-Vir}
S=({\mathrm{Vir}} \oplus 
{\cal S}^F\left({\cal V}_{\mu/F}\right)_{\mathrm{red}}) \oplus
{\cal V}_{\mu/F}
\eeq

\noindent
with  obvious notation (see (\ref{eq:F-g}). Of course this construction is 
different from the
previous one (\ref{eq:falg}) since ${\cal V}_{\mu/F}$ is a highest weight
representation of Vir.

\subsection{The construction of $V(g)$}

In this section, we will construct an infinite dimensional Lie algebra 
$V(g)$ which contains the Lie algebra $g$ in the same way as Vir contains
$sl(2)$.

Let $g$ be a semi-simple complex Lie algebra, let $h$ be a Cartan sub-algebra,
let $\Phi_+$ be the  positive roots and let 
$(\alpha_{(1)},\cdots,\alpha_{(r)})$
be the positive  simple roots. \\
We consider the vector space $V$  generated by 

\beqa
\label{eq:gene}
V=\Big<L_0^{\alpha_i},~~ L^\alpha_{n}:~~
i=1,\cdots,r, ~~\alpha \in \Phi_+,~~ n \in \ZZ^* \Big>
\eeqa

\noindent
satisfying the commutation relations:  

\begin{enumerate}
\item

\begin{enumerate}
\item for positive simple roots $\alpha_{(1)},\cdots,\alpha_{(r)}$,
\beqa
\label{eq:vir-g1}
\left[L_0^{\alpha_i}, L_0^{\alpha_j} \right]=0.
\eeqa

\item
for $n >0$,

\beqa
\label{eq:vir-g2}
\left[{L_0^{\alpha_i} \over- n}, {L^\alpha_{\pm n} 
\over \pm n} \right] 
&=& \pm{ \alpha_{(i)} . \alpha \over \alpha_{(i)}^2}{L^\alpha_{\pm n} \over 
\pm n},~  i=1,\cdots,r,  \alpha \in \Phi_+.  \nonumber \\
\left[{L^\alpha_n\over n},{L^\beta_n \over n}\right] &=& \left \{
\begin{array}{ll}
\epsilon\{\alpha,\beta\} {L^{\alpha+\beta}_n\over n}& {\mathrm {~~if~~}}
\alpha + \beta \in \Phi_+ \cr
0& {\mathrm {~~otherwise }}.
\end{array}
\right. \nonumber \\
\left[{L^\alpha_n \over n}{,L^\beta_{-n} \over - n} \right] &=& \left \{
\begin{array}{ll}
\epsilon(\alpha,-\beta){ L_n^{\alpha-\beta} \over n}& {\mathrm {~~if~~}}  
\alpha - \beta \in \Phi_+ \cr
\epsilon(\alpha,-\beta) {L_{-n}^{-\alpha+\beta} \over -n}& {\mathrm {~~if~~}}  
-\alpha + \beta \in \Phi_+ \cr
2 {L^\alpha_0 \over -n} & {\mathrm {~~if~~} } \alpha=\beta. \cr
\end{array}
\right. \\
\left[{L^\alpha_{-n} \over -n},{L^\beta_{-n}\over -n}\right] &=& \left \{
\begin{array}{ll}
\epsilon\{-\alpha,-\beta\} {L^{\alpha+\beta}_{-n}\over -n}& {\mathrm {~~if~~}}
\alpha + \beta \in \Phi_+ \cr
0& {\mathrm {~~otherwise. }}
\end{array}
\right. \nonumber
\eeqa

\noindent
Thus $g_n=\Big<L_0^{\alpha_i}, L_{\pm n}^{\alpha}: 
\alpha \in \Phi_+, i=1,\cdots,r\Big>$
is a Lie algebra isomorphic to $g$,   with Cartan sub-algebra 
$\Big<L_0^{\alpha_i}, i=1,\cdots,r \Big>$, roots $\Phi$, simple
positive roots $\alpha_{(1)},\cdots, \alpha_{(r)}$ and root spaces 
$\Big<L^\alpha_{\pm n} \Big>$.
The isomorphism with (\ref{eq:lie}) is: 

\beqa
\label{eq:gn}
{L_0^{\alpha_i} \over- n} &\Leftrightarrow& {\alpha_{(i)} . 
H \over \alpha_{(i)}^2} \\ \nonumber 
{{L^\alpha_n} \over n} &\Leftrightarrow& E^\alpha \\
{{L^\alpha_{-n}} \over -n} &\Leftrightarrow& E^{-\alpha} \nonumber.
\eeqa

It is important to emphasize that the Cartan sub-algebra of each $g_n$ is
independent of $n$. 
Consequently the relation $[L^\alpha_1,L^\alpha_{-1}]=2 L^\alpha_0$
defines with no ambiguity $L^\alpha_0$ when $\alpha$ is not a simple root
(it is just a linear combination of the $L^{\alpha_i}_0$ where
$\alpha_{(i)}, i=1,\cdots,r$ are the simple roots).
\end{enumerate}
\item

With the notation  $L^\alpha_0= {1 \over 2n} [L^\alpha_n,L^\alpha_{-n}]$ 
(this is independent of $n>0$ by (\ref{eq:vir-g2})) for $\alpha \in \Phi_+$:

\beqa
\label{eq:vir-g3}
\big[L_n^\alpha,L_m^\alpha\big]=(n-m)L_{n+m}^\alpha.
\eeqa
 
\end{enumerate}








\noindent

The relations (\ref{eq:vir-g1}), (\ref{eq:vir-g2}) and (\ref{eq:vir-g3}) do 
not specify all
commutators $[L^{\alpha}_n,L^\beta_m]$ ({\it e.g.}
$[L^\alpha_1, L^\beta_2]$ if $\alpha \ne  \beta$).
In order to obtain a Lie algebra from (\ref{eq:vir-g1}), (\ref{eq:vir-g2})
 and (\ref{eq:vir-g3}) 
we define 

\beqa
\label{eq:nivvg}
V(g) = T\Big(V\Big)/{\cal I},
\eeqa

\noindent
where $T\Big(V\Big)$  is
the tensorial algebra on 
$V$
and ${\cal I}$ is the two-sided ideal generated by the relations 
(\ref{eq:vir-g1}), (\ref{eq:vir-g2}) and (\ref{eq:vir-g3}). 
$V(g)$ is an associative algebra
but  we will consider it as a Lie algebra for the induced Lie bracket 
({\it i.e.} commutator). The universal property of $V(g)$ is the following:
any linear map $f:V
\rightarrow {\mathrm{End}}(H)$ such that the $f(L_n^\alpha)$ satisfy the 
relations  (\ref{eq:vir-g1}), (\ref{eq:vir-g2}) and  (\ref{eq:vir-g3}) extends
 to a unique Lie algebra 
homomorphism $\tilde f:V(g) \rightarrow{\mathrm{End}}(H)$. The relations
(\ref{eq:vir-g1}), (\ref{eq:vir-g2}) and (\ref{eq:vir-g3}) can be arranged in
the following diagram (for $g=su(3)$)
   
\begin{figure}[!h]
 \epsfysize =10.cm
$$\epsffile{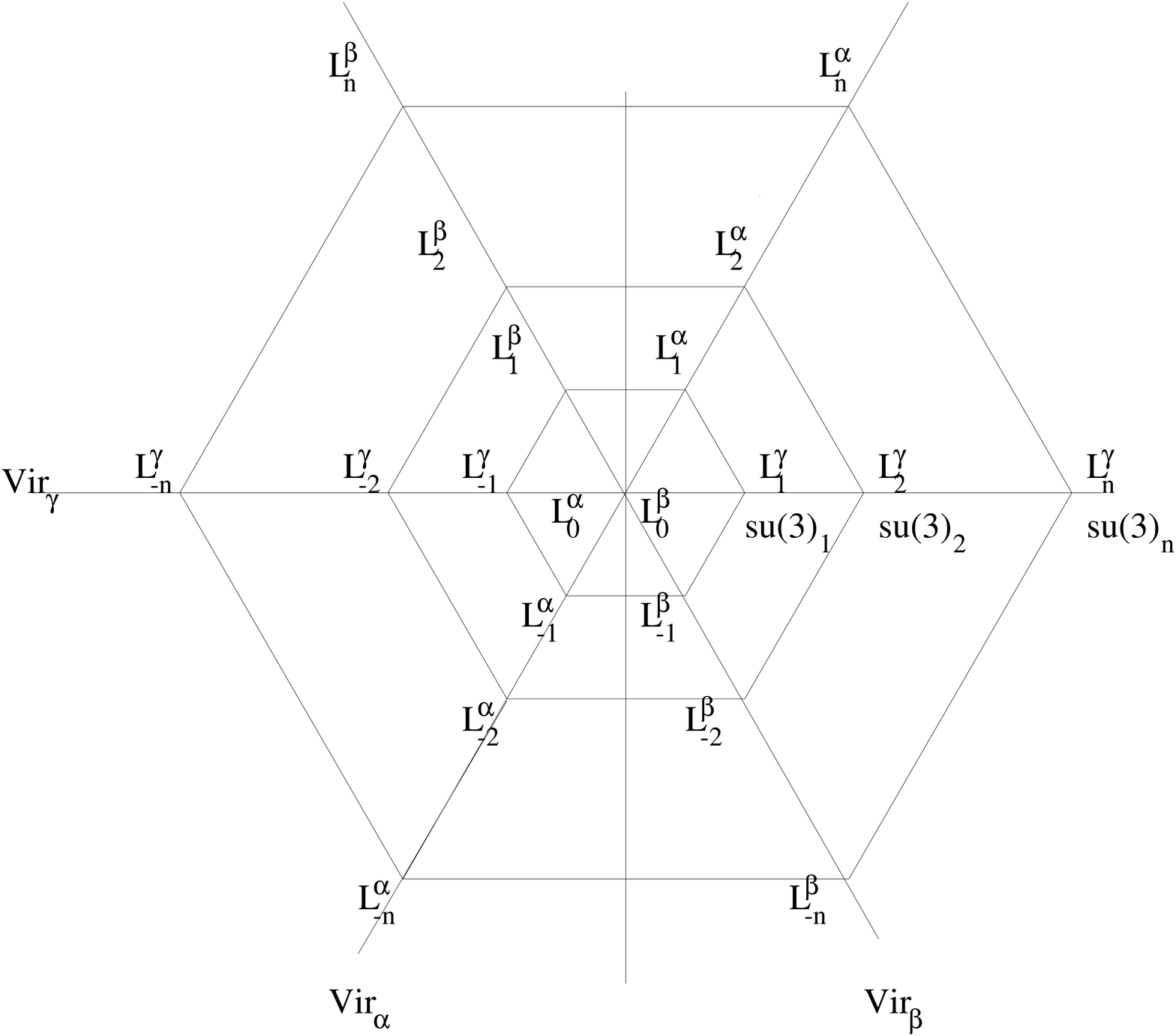}$$
\caption{ Digram for the primitive generators for the 
$V(su(3))$ algebra. $\alpha, \beta$ are the primitive roots and
$\gamma$ is the third positive root.}
\end{figure}       

Of course similar diagrams can be constructed for higher rank Lie algebra,
and the results can be easily extend to all Lie algebras $g$.
This diagram has  a concentric and radial structure. 
The concentric symmetry is just the manifestation that to any positive $n$ 
can be associated an  algebra isomorphic to $g$ (see (\ref{eq:vir-g2})).
The radial symmetries extend the $sl(2)_\alpha$ (generated by 
$(L^\alpha_0, L^\alpha_{\pm 1})$)
algebra to  ${\mathrm Vir}_\alpha$ a Virasoro algebra 
(see (\ref{eq:vir-g3})). 
Finally, composing concentric and radial symmetries through  Jacobi 
identities generates extra   symmetries {\it i.e.} the secondary, ternary ... 
generators. 
The generators $L_n^{\alpha_i}$ with $n>0$ and $\alpha_i$ a 
simple root are  called the fundamental generators (or primary) of $V(g)$ 
in the sense that all the others can be obtained from them by taking
commutators.
The primary generator $L^\alpha_n$  is associated to the root $n \alpha$.
Calculating secondary, ternary, ... generators through Jacobi identities
induces generators associated to a root $\sum n_i \alpha_i$. 
For instance if $\alpha + \beta \in \Phi$, the root
associated to $[L^\alpha_n,L^\beta_m]$ is $n \alpha + m \beta$. 
The Lie algebra $V(g)$ is clearly not a Kac-Moody algebra, but according to
Kac classification of Lie algebras
\cite{k}, $V(g)$ is an infinite dimensional Lie algebra with a non-polynomial 
degeneracy of the  roots. 
Let us stress again that, in such diagrams,
the Cartan sub-algebra is at the origine of the diagram, and the 
bracket between two generators is defined {\it only} between
generators on the same circle, or the same radius.

The decomposition 
\beqa
\label{eq:vect}
V=\Big<L_0^{\alpha_i}: i=1,\cdots,r\Big> \oplus \Big<L_n^\alpha: \alpha \in 
\Phi_+, n > 0 \Big>
\oplus \Big<L_{-n}^\alpha: \alpha \in \Phi_+, n > 0 \Big>
\eeqa

\noindent
should be thought of as a decomposition of $V$ into a ``Cartan`` sub-algebra 
and 
positive and negative ``root'' spaces. From this point of view, one expects to
be able to construct representations of $V(g)$ from ``weights'', {\it i.e.}
from linear forms on the Cartan sub-algebra 
$\Big<L_0^{\alpha_i}: i=1,\cdots,r\Big>$.
Indeed, if $\mu \in \Big<L_0^{\alpha_i}: i=1,\cdots,r\Big>^*$ 
(the dual space of $\Big<L_0^{\alpha_i}: i=1,\cdots,r\Big>$) 
is a weight then 
${\cal U}\Big( V(g)\Big)/I_\mu$ is a representation of $V(g)$ where $I$ is the
 two-sided ideal in ${\cal U} \Big(V(g)\Big)$ generated by the 
$L^{\alpha_i}_0 - \mu(L^{\alpha_i}_0) {\mathbf 1}~
(i=1,\cdots,r)$ and the $L^\alpha_n ~(n > 0,  \alpha \in \Phi_+)$ 
(see section 3.2). As in section 3.2 

\beq
\label{eq:F-V(g)}
S=(V(g) \oplus {\cal S}^F\left({\cal V}_{\mu/F}\right)_{\mathrm{red}}) \oplus
{\cal V}_{\mu/F}
\eeq

\noindent
is an $F-$Lie algebra (with the obvious notation).

\subsection{Examples for $so(6)$}
The Verma module construction in the previous sub-section gives many
examples of $F-$Lie algebras. However, the representations of $V(g)$
obtained  are all highest  weight representations. In this
section, in the spirit of section 4.1, we will construct explicitly
non-highest weight representations of $V(so(6))$ and the corresponding
$F-$Lie algebras.

First of all,  we introduce
$H_1, H_2, H_3$ the generators of the Cartan sub-algebra. We denote $\pm e^i$
the eigenvalues of $H_i$,
and 
$\Phi=\left\{\alpha_1=e^1-e^2, \alpha_2=e^2-e^3,\right.$ 
$\left.\alpha_3 = e^2+e^3, 
\beta_1=e^1+e^2,\right.$ $\left.\beta_2=e^1-e^3, \beta_3=e^1+e^3 \right \}$ 
the positive
roots, where $\alpha_1,
\alpha_2$ and $\alpha_3$ are the simple roots.
Then we introduce the six $sl(2):
\left\{E^{\pm \phi}, 
{\phi.H,  \over \phi^2}, ~ \phi \in \Phi\right \}$. 
The three $sl(2)$ associated to the  primitive roots are then generated by

\beqa
\label{eq:sl2}
\begin{array}{ll}
sl(2)_1:&\Big\{ E^{\pm \alpha_1}, {1 \over 2} \left(H_1-H_2\right) \Big \} \cr
sl(2)_1:&\Big\{ E^{\pm \alpha_2}, {1 \over 2} \left(H_2-H_3\right) \Big \} \cr
sl(2)_3:&\Big\{ E^{\pm \alpha_3}, {1 \over 2} \left(H_2+H_3\right) \Big \}.
\end{array}
\eeqa

Secondly,  we need to understand, using the results of section 4.1, how one
can extend the representations  (\ref{eq:repp}) of $so(1,2)$  to 
representations (\ref{eq:repv}) of Vir.
For that purpose, we introduce
commuting variables belonging to the $so(1,2)$ representation: 
\begin{itemize}
\item if the representation is a spin-$s$ representation we define 
$\big\{x_{-s},\cdots,x_s\big\}=\D_{-s}$;
\item 
for infinite dimensional representation bounded from below/above we consider
$\D_{\lambda}^\pm=\big\{x_p, p \in \pm(\NN - \lambda)\big\}$.
\end{itemize}

Next, using the results of section 4.1, we can extend any representations
$\D$ of  $so(1,2)$ to a representation $\hat\D$ of 
the  Virasoro algebra. Explicitly:

\begin{itemize}
\item For $\hat \D_{-s}$, one can easily observe that $x_{-ns}$,
for any positive $n$ when $s$ integer and for odd positive  $n$ when $s$ 
half-integer,
is a primitive
vector for the $sl(2)-$ sub-algebra spanned by $\left\{L_{\pm n}/n, 
L_0/n\right\}$
so $\D_{-s}^n=\left\{x_{-ns},x_{-(n-1)s},\cdots, x_{ns} \right\}$ is 
isomorphic to
a $\D_{-s}$ representation;
\item in the case of infinite dimensional representations we have similar
results if $\lambda$ is a rational number. In the case of interest $\lambda=
1/F$, we have primitive vector only when $n=\pm(pF+1)$.

\end{itemize}
 
To conclude with the construction of the representation of Vir,
we just have to introduce 
$p_\ell$ the conjugate momentum of 
$x_k, [p_{-\ell},x_k] = \delta_{k \ell}$)

\begin{itemize}
\item For $\hat \D_{-s}=\big\{x_n, n \in \ZZ + s \big\}$ the generators of 
the Virasoro algebra are 
$L_n= \sum \limits_{k \in \ZZ} \big(ns-m) x_{n+m} p_{-m}$;
\item and for $\hat \D^\pm_\lambda=\big\{x_p, p \in \ZZ \mp \lambda \big\}$:
$L_n= \sum \limits_{k \in \ZZ \mp \lambda} \big(n \lambda - k) x_{n+k} p_{-k}$.
\end{itemize}
 
In conformal field theory these expressions for the
generators of the Virasoro algebra are known \cite{pol}. 
The $L_n$ are the modes
of the stress-energy tensor (for the conformal or superconformal ghosts)
associated to conjugate conformal fields
of conformal weight $1+ \lambda$ (the $x$'s) and $-\lambda$ (the $p$'s)
But it is important to emphasize that
in contrast with the case of
conformal field theory, here we do not have any normal ordering
prescription and consequently no central charge.

\subsubsection*{The spin representation of so(6)}

The spinorial representation of chirality $+$ is obtained from the
dominant weight $\mu_+= \mu_{(3)}={1 \over 2} (e^1+e^2+e^3)$ 
and the highest weight state 
is  $\left| {1 \over 2},{1 \over 2},{1 \over 2}\right>$. Then by the action of
$E^{-\alpha_i}$ we get the whole representation

$${\cal D}_{\mu_+} = \left \{
\left| {1 \over 2},{1 \over 2},{1 \over 2} \right>,
\left| {1 \over 2},-{1 \over 2},-{1 \over 2} \right>,                         
\left| -{1 \over 2},{1 \over 2},-{1 \over 2} \right>,
\left| -{1 \over 2},-{1 \over 2},{1 \over 2} \right> \right \}. $$

\noindent
(In $\left|a_1,a_2,a_3 \right>, a_i$ is the eigenvalue of
$H_i$).

Now, it is easy to see that with respect to the three $sl(2)$ (noted $sl(2)_i$)
we have three spinorial representations (for each $sl(2)_i$, we have 
$\Big(\D_{-1/2}\Big)_i \subset \D_{\mu_+})$

\beqa
\label{eq:spin}
\begin{array}{ll}
sl(2)_1~:&\Big(D_{-1/2}\Big)_1= \left\{
x^{(1)}_{1/2} \equiv\left| {1 \over 2},-{1 \over 2},-{1 \over 2} \right>,
x^{(1)}_{-1/2} \equiv \left| -{1 \over 2},{1 \over 2},-{1 \over 2} \right> 
\right\}\cr
sl(2)_2~:&\Big(D_{-1/2}\Big)_2=\left\{ 
x^{(2)}_{1/2}\equiv\left| -{1 \over 2},{1 \over 2},-{1 \over 2} \right>,
x^{(2)}_{-1/2}\equiv \left| -{1 \over 2},-{1 \over 2},{1 \over 2} \right> 
\right\}\cr
sl(2)_3~:&\Big(D_{-1/2}\Big)_3=\left\{
x^{(3)}_{1/2}\equiv \left| {1 \over 2},{1 \over 2},{1 \over 2} \right>,
x^{(3)}_{-1/2}\equiv\left| {1 \over 2},-{1 \over 2},-{1 \over 2} \right>
\right\}
\end{array}
\eeqa

\noindent
where in $x^{(i)}_m$, $i$ indicates to which $sl(2)$ the states belong  and
$m$ is the eigenvalue of ${1 \over 2} \alpha_i.H$. Now, having introduced these
variables, we can straightforwardly write the generators associated to the
three $sl(2)$. For $sl(2)_i$, we have (at this point we already take the 
normalizations appropriate for  the $V(so(6))$ generalization)

\beqa
\label{eq:spin2}
E^{-\alpha_i}=J^{(i)}_-&=&-x^{(i)}_{-1/2} p^{(i)}_{-1/2} \nonumber \\
{1 \over 2} \alpha_i.H=J_0^{(i)}&=&{1 \over 2} \left( 
x^{(i)}_{-1/2} p^{(i)}_{1/2}-
x^{(i)}_{1/2} p^{(i)}_{-1/2} \right) \\
E^{\alpha_i}=J^{(i)}_+&=&x^{(i)}_{1/2} p^{(i)}_{1/2} \nonumber
\eeqa

\noindent
with $p$ the conjugate momentum of $x$ ($[p_{-\ell},x_k]=\delta_{\ell k}$).

Of course all the $x$  variables are not independent and the following 
identifications have to be made

\beqa     
\label{eq:ident1}
\begin{array}{ll}
x^{(1)}_{-1/2}=x^{(2)}_{1/2}&p^{(1)}_{1/2}=p^{(2)}_{-1/2} \cr
x^{(1)}_{1/2}=x^{(3)}_{-1/2}&p^{(1)}_{-1/2}=p^{(3)}_{1/2},
\end{array}
\eeqa

\noindent
leading to dependent $sl(2)$.

Now, the next point is to introduce new variables which allow us to define
the $\hat {\cal D}_{\mu_+}$ representation of $V(so(6))$. In other 
words, we need
to extend the $J^{(i)}_\pm, J^{(i)}_0$ operators to $L^{(i)}_n, n \in \ZZ$. 
This can be done,  by
considering the three $sl(2)_{2p+1}$ span by $L_{\pm n}^{(i)}/n, L_0^{(i)}/n$ 
with
$n=2p+1, p \in \NN$. We extend the $\Big(\D_{-1/2}\Big)_i$ rep. of $sl(2)_i$
into a rep. of ${\mathrm Vir}_i$. For that we introduce a highest weight state 
$\left|\mu_p\right>= \left|{2p+1 \over 2},{2p+1 \over 2}, 
{2p+1 \over 2} \right>$ (corresponding to a primitive vector
of the $sl(2)_{2p+1}$),
 and construct explicitly the induced spinorial 
representation for $so(6)$:

\beqa
\label{eq:spinind}
{\cal D}^p_{\mu_+} = &&\left \{
\left| {2p+1\over 2},{2p+1 \over 2},{2p+1 \over 2} \right>,
\left| {2p+1 \over 2},-{2p+1 \over 2},-{2p+1 \over 2} \right>,  \right.
 \\
&&\left| -{2p+1 \over 2},{2p+1 \over 2},-{2p+1 \over 2} \right>, 
\left.\left| -{2p+1 \over 2},-{2p+1 \over 2},{2p+1 \over 2} \right>
 \right\}. 
\nonumber
\eeqa

\noindent
As previously, we interpret the various states as spinorial multiplets
of the three $sl(2)$, and we define $6$ variables $x^{(i)}_{\pm(2p+1)}$.
Now arguing that the $x^{(i)}_r, r \in \ZZ + 1/2$ span a 
$\left(\hat {\cal D}_{-1/2}\right)_{i}$
representation of the Virasoro algebra, we set  

\beqa
\label{eq:viri}
L_n^{(i)} = \sum \limits_{r \in \ZZ +1/2} ({n \over 2} - r)
x^{(i)}_{n+r} p^{(i)}_{-r}.
\eeqa

\noindent
To conclude the construction of the 
$\hat {\cal D}_{\mu_+}=\oplus_p \D_{\mu,+}^p$ representation,
we make the same identifications as in (\ref{eq:ident1})

\beqa
\label{eq:ident2}
\begin{array}{ll}
x^{(1)}_{-(2p+1)/2}=x^{(2)}_{(2p+1)/2}&
p^{(1)}_{(2p+1)/2}=p^{(2)}_{-(2p+1)/2} \cr
x^{(1)}_{(2p+1)/2}=x^{(3)}_{-(2p+1)/2}&
p^{(1)}_{-(2p+1)/2}=p^{(3)}_{(2p+1)/2}
\end{array}
\eeqa

This specified the $\hat \D_{\mu_+}$ representation. The remaining commutators
of $V(so(6))$ can then be calculated explicitly.

\subsubsection*{The Vector representation}
For the vector representation the dominant weight is $\mu_v=\mu_{(1)}=e^1$.
To construct $\hat {\cal D}_{\mu_v}$ from ${\cal} D_{\mu_v}$,
we proceed along the same lines as for the spin representation, so we
will be less explicit in our construction. 
Constructing $\D_{\mu_v}$, we have for each $sl(2)_i$ the following 
decomposition $\D_{\mu_v} \supset \Big(\D_{-1/2}\Big) \oplus 
\Big(\D^\prime_{-1/2}\Big)$. 
We define  highest weights
associated to the three $sl(2)$ generated by $(L_{\pm}^{(i)}/n, L_0^{(i))}/n)$
with $n=2p+1$. This induce a vector representation $\D^p_{\mu_v}$:

\beqa
\label{eq:revvect}
{\cal D}^p_{\mu_v} =\left\{\left|\pm(2p+1),0,0\right>, \left|0,\pm(2p+1),0
\right>, \left|0,0\pm(2p+1)\right> \right \}.
\eeqa

\noindent
We then identify six spinorial representations of the primitive $sl(2)$:

\beqa
\label{eq:spinv}
\begin{array}{ll}
sl(2)_1~:&\Big(\D^p_{-1/2}\Big)_1=
\left\{x^{(1)}_{(2p+1)/2}\equiv \left|2p+1,0,0\right>,
      x^{(1)}_{-(2p+1)/2}\equiv \left|0,2p+1,0\right> \right\}\cr
&\Big(\D^{\prime p}_{-1/2}\Big)_1=
\left\{ x^{\prime~(1)}_{(2p+1)/2}\equiv \left|0,-(2p+1),0\right>, 
        x^{\prime~(1)}_{-(2p+1)/2}\equiv \left|-(2p+1),0,0\right> \right\}  \cr
sl(2)_2~:&\Big(\D^p_{-1/2}\Big)_2=
\left\{x^{(2)}_{(2p+1)/2}\equiv \left|0,2p+1,0\right>,
      x^{(2)}_{-(2p+1)/2} \equiv \left|0,0,2p+1\right> \right\} \cr
&\Big(\D^{\prime p}_{-1/2}\Big)_2= 
\left\{x^{\prime~(2)}_{(2p+1)/2}\equiv \left|0,0,-(2p+1)\right>, 
       x^{\prime~(2)}_{-(2p+1)/2}\equiv \left|0,-(2p+1),0\right>\right\}  \cr
sl(2)_3~:&\Big(\D^p_{-1/2}\Big)_3=
\left\{x^{(3)}_{(2p+1)/2}\equiv \left|0,2p+1,0\right>,
      x^{(3)}_{-(2p+1)/2}\equiv \left|0,0,-(2p+1)\right>\right\} \cr
&\Big(\D^{\prime p}_{-1/2}\Big)_3= 
\left\{x^{\prime~(3)}_{(2p+1)/2}\equiv \left|0,0,(2p+1)\right>,
       x^{\prime~(3)}_{-(2p+1)/2}\equiv \left|0,-(2p+1),0\right> \right\}  
\end{array}
\eeqa

\noindent
The appropriate identifications can be read off from (\ref{eq:spinv})

\beqa
\label{eq:identv}
\begin{array}{ll}
x^{(1)}_{-(2p+1)/2}=x^{(2)}_{(2p+1)/2}=x^{(3)}_{(2p+1)/2}&
p^{(1)}_{(2p+1)/2}=p^{(2)}_{-(2p+1)/2}=p^{(3)}_{-(2p+1)/2} \cr
x^{\prime~(1)}_{(2p+1)/2}=x^{\prime~(2)}_{-(2p+1)/2}=x^{\prime~(3)}_{-(2p+1)/2}
&
p^{\prime~(1)}_{-(2p+1)/2}=p^{\prime~(2)}_{(2p+1)/2}=p^{\prime~(3)}_{(2p+1)/2}
\cr
x^{(2)}_{-(2p+1)/2}= x^{\prime (3)}_{(2p+1)/2}&
p^{(2)}_{(2p+1)/2}= p^{\prime (3)}_{-(2p+1)/2} \cr
x^{\prime (2)}_{(2p+1)/2}=x^{(3)}_{-(2p+1)/2}&
p^{\prime (2)}_{-(2p+1)/2}=p^{(3)}_{(2p+1)/2}
\end{array}
\eeqa 

\noindent
So, finally we obtain the explicit expression for the primitive (associate
to the simple roots of $so(6)$) generators of $V(so(6))$
for the $\hat \D_{\mu_v} = \oplus_p \D_{\mu_v}^p$ representation.

\beqa
\label{eq:virvec}
L_n^{(i)}= \sum \limits_{r \in \ZZ +1/2} (n- {r \over 2})
\left(x^{(i)}_{n+r} p^{(i)}_{-r} + x^{\prime~(i)}_{n+r} p^{\prime~(i)}_{-r}
\right).
\eeqa

\subsubsection*{The $\D^+_{{\mu_+ \over F}}$ representation of $so(6)$}

In this example, we show how one can obtain an explicit realization of
the $\hat \D^+_{{\mu_+ \over F}}$ representation without giving any 
differential
realization. Starting with the highest weight state
$\left| {\mu_+ \over F} \right>= 
\left|{1 \over 2F},{1 \over 2F},{1 \over 2F}\right>$
we construct the $\D^+_{{\mu_+ \over F}}$ representation. 
($\hat \D^-_{{\mu_+ \over F}}$ would have been obtained from 
$\left|-{1 \over 2F},-{1 \over 2F},{1 \over 2F}\right>$).
The interesting point
with such a representation is that the states of the third $sl(2)$ belong
to $\D_{1/F}^+$ and of  the first and second $sl(2)$ to finite dimensional
representations. Now, to obtain the whole representation 
$\hat \D_{{\mu_+ \over F}}$ it is enough to find the primitive vector 
associated
to $(L^{(i)}_{\pm (2Fp+1)}/(2Fp+1),L^{(i)}_0/(2Fp+1))$
(there are no other states annihilated by some $L_n$,
see (\ref{eq:repv}) and (\ref{eq:viract})). This vector,
$\left|{1+2Fp \over 2F},{1+2Fp \over 2F},{1+2Fp \over 2F}\right>$ induces a 
$\D^+_{{\mu_+ \over F}}$
representation. With the analogous identifications  as for the vectorial
and spinorial representations we get $\hat \D_{{\mu_+ \over F}}$.

\subsubsection*{Application to $F-$Lie algebras with $so(6)$}

Having constructed the representations of $V(so(6))$, one can easily
prove that $\hat \D^+_{\mu_+}$ is included in 
${\cal S}^F\left(\hat \D^+_{\mu_+ \over F} \right)_{\mathrm red}$
(${\cal S}^F\left(\hat \D^+_{\mu_+ \over F} \right)_{\mathrm red}$
is defined as in section 3.2). 
We just have to notice that 

$$\hat \mu_F \left( {\cal S}^F \left(
\left|p_1+{1 \over 2F},p_1+{1 \over 2F},p_1+{1 \over 2F} \right>, \cdots, 
\left|p_F+{1 \over 2F},p_F+{1 \over 2F},p_F+{1 \over 2F} \right> \right) 
\right)$$

\noindent
 with 
$\sum p_i = p$ is a primitive vector which induces the spinorial 
representations
(\ref{eq:spinind}) obtained with $\left| \mu_p \right>$ in $\D_{\mu_+}$.
This leads to the $F-$Lie algebra 
$\Big(V(so(6)) \oplus  (\D^+_{{\mu_+ \over F}})_{\mathrm red} 
\Big)
\oplus \hat  \D^+_{{\mu_+ \over F}}$ \\
\vskip.5truecm
To conclude, the results of this sub-section probably extend, along
the same lines, to any Lie algebra $g$.

\mysection{Conclusion}

Supersymmetry is a well established mathematical structure ($\ZZ_2$ 
graded algebras)
and beyond its purely formal aspects it has found a wide range of applications
in field theories and particle physics. 
In this article, we have defined a 
mathematical structure  generalizing the concept
of super-Lie algebras ($F-$Lie algebras). These algebras seems to be
appropriate if one wants  to generalize supersymmetry in the sense of 
$F^{{\mathrm th}}-$roots
of representations. Indeed, we have shown that, 
within the framework of $F-$Lie
algebras, it is possible to take the 
$F^{{\mathrm th}}-$root of {\it any representation
of any (complex semi-simple) Lie algebra}. 
In addition, $F-$Lie algebras naturally lead to 
the infinite
dimensional Lie algebra $V(g)$ containing $g$ as a sub-algebra. For the
special case  $g=sl(2,\RR)$, $V(g)$  is the centerless Virasoro algebra.
As a consequence, one may wonder  whether or not central extensions
of $V(g)$ exist. Furthermore, one has a geometrical interpretation
of the Virasoro algebra (as vector fields on the circle) so is there a
similar interpretation for $V(g)$ ?

Unitary representations of FSUSY, for $g=so(1,2)$ have also been constructed.
It has also been checked that it is
a symmetry  acting on  relativistic anyons \cite{fsusy3d}. 
In the same way, since the Lorentz group in higher dimensions is just 
$SO(1,d-1)$, what is the interpretation of FSUSY for $g=so(1,d-1)$ when the 
$F-$Lie algebra induces  the $F-$root of the spin or the vector 
representations ?

\vskip1truecm
{\bf Acknowledgements:} We are  grateful to A. Neveu and 
J. Thierry-Mieg for  useful  discussions and critical comments.


\end{document}